\documentclass[12pt]{article}
\usepackage{lscape} 
\usepackage{amsthm}
\usepackage{amsmath}
\usepackage{setspace}
\usepackage{epsf}
\usepackage{csquotes} 
\usepackage{rotating}
\usepackage{fancyhdr}
\usepackage{graphicx}   
\usepackage{ytableau}
\usepackage{enumitem}
\usepackage{placeins}
\usepackage[labelfont=bf]{caption}
\usepackage{subfigure}
\usepackage{amssymb}
\usepackage{verbatim}
\usepackage{bbold}
\usepackage[export]{adjustbox}
\usepackage{cancel}
\usepackage{cite}
\usepackage{amsfonts}
\usepackage{tikz}
\usepackage[authoryear,longnamesfirst]{natbib}

\newcommand{\micron}{$\mu$m }

\def\tsc#1{\csdef{#1}{\textsc{\lowercase{#1}}\xspace}}
\tsc{WGM}
\tsc{QE}
\usepackage{color,xcolor,soul}

\begin{document}
\include{epsf} 
\begin{titlepage}
\begin{flushright}
FTUV-25-0804.1851
\end{flushright}

\begin{center}
{\Large \bf 
  Gauge theory approach to describe 
  \\[0.1cm]
  ice crystals habit evolution in ice clouds \\[0.1cm]
}
\end{center}

\par \vspace{2mm}
\begin{center}
  {\bf  Gianluca Di Natale${}^{(a)}$},  {\bf Francesco Pio De Cosmo${}^{(b)}$}
  and {\bf Leandro Cieri${}^{(c)}$}\\

\vspace{5mm}

${}^{(a)}$ 
Consiglio Nazionale delle Ricerche, National Institute of Optics, Via Madonna del Piano, 10, Sesto Fiorentino, Firenze, Italy.\\\vspace{1mm}

${}^{(b)}$ 
Consiglio Nazionale delle Ricerche, Institute for Application of Calculous, Via Madonna del Piano, 10, Sesto Fiorentino, Firenze, Italy. \\\vspace{1mm}

${}^{(c)}$ 
Instituto de F\'isica Corpuscular (IFIC), CSIC - Universitat de Val\`encia, Parc Cient\'ific, E-46980 Paterna, Valencia, Spain\\\vspace{1mm}
\end{center}

\vspace{1.5cm}

\par \vspace{2mm}
\begin{center} {\large \bf Abstract} \end{center}
\begin{quote}
\pretolerance 10000
Ice clouds, particularly cirrus clouds, significantly influence Earth's radiative balance but remain poorly characterized in current climate models. A major uncertainty arises from the variability of their microphysical properties, especially the evolution of ice crystal habits under depositional growth. We propose a heuristic method to describe habit evolution based on four fundamental shapes identified in the literature and from in situ observations: droxtals, plates, columns, and rosettes. These represent the primary forms that are relevant under depositional growth, excluding aggregation. In this study, we employ a non-Abelian gauge theory within a field-theoretical framework, imposing an SU(2) $\otimes$ U(1) symmetry on the fields associated with each habit probability growth. This symmetry enables the derivation of a modified system of coupled Fokker–Planck equations, capturing the stochastic growth dynamics of ice crystals while incorporating phenomenological mutual influences among habits. This framework outlines a novel theoretical direction for integrating symmetry principles and field-theoretical tools into the modelling of habit dynamics in ice clouds.
\end{quote}

\vspace*{\fill}
\vspace*{2.5cm}

\begin{flushleft}
August 2025
\end{flushleft}
\end{titlepage}

\setcounter{footnote}{1}
\renewcommand{\thefootnote}{\fnsymbol{footnote}}

\section{Introduction}\label{Intro}
Nowadays, the issue of climate change is one of the most urgent problems afflicting our planet. Improving the performance of climate models is closely related to accurately parameterising clouds, particularly ice clouds, since these are still very uncertain and not fully understood, representing a major source of error in such models. In particular, it has been shown that cirrus clouds --- a type of ice cloud --- have a strong impact on the Earth's energy balance \cite{Lynch1996}. In fact, depending on their optical and micro-physical properties, they can either cool or warm the planet \cite{Baran2009}. For example, some studies have shown that crystals larger than 24 \micron exert positive feedback, while smaller crystals lead to negative feedback \cite{Stephens1994}. This results in different cloud forcing (CF),  which is defined as the difference in emitted radiance between cloudy and clear-sky conditions \cite{Intrieri2002}, and can affect the warming of the planet's surface.

Over the past few decades, significant progress has been made in modelling ice cloud properties and developing databases of the single scattering properties of ice crystals, as described in \cite{Fu1993,Yang2013,Baran2009}. These databases were obtained using various calculation techniques, including Mie theory, the finite difference time domain (FDTD) method, the discrete dipole approximation and geometrical optics. These properties enable the simulation of the interaction between ice crystals and radiation, taking into account their highly non-spherical shapes, and the simulation of radiative transfer in the atmosphere. Unfortunately, these properties and models have not yet been fully validated. Nevertheless, there is still a lack of measurements, particularly in the far infrared (FIR) portion of the spectrum, from which about $50\%$ of the total flux emitted by the atmosphere originates \cite{brindley1998}. Furthermore, the fact that ice clouds permanently cover a large portion of the planet's surface (in particular, cirrus clouds — the coldest and highest type of ice cloud — cover 30$\%$, reaching 70$\%$ in the tropics) highlights the importance of accurate ice cloud modelling. Near-future missions are planned to explore the far-infrared portion of the spectrum for the first time up to a wavelength of 100 \micron and to study the properties of ice clouds, such as the \textit{Far Infrared Outgoing Radiation for Understanding and Monitoring} (FORUM) mission, which was selected by the European Space Agency (ESA) \cite{Palchetti2016_proc,Palchetti2020} as the next Earth Explorer-9.

The various habits of ice crystals exhibit different behavior with regard to radiation absorption and scattering. This affects radiative transfer through the atmospheric layers, meaning that assuming a specific crystal habit distribution in cloud modelling can lead to large errors in calculating CF and outgoing longwave radiation (OLR). These errors consequently increase those affecting the performance of climate model predictions. Although we need to assess the habit distribution in ice clouds as accurately as possible, it is unfortunately not easy to assume these distributions since in situ and remotely (from satellites, aircraft or ground stations) measurements aiming for this purpose are still scarce.

\begin{figure}
    \centering
    \includegraphics[width=0.2\linewidth]{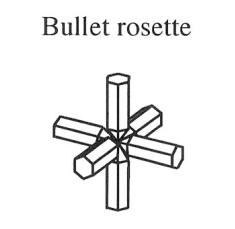}
     \includegraphics[width=0.2\linewidth]{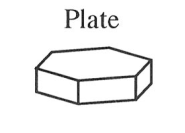}
      \includegraphics[width=0.2\linewidth]{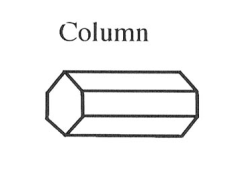}
       \includegraphics[width=0.2\linewidth]{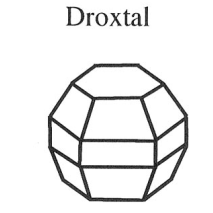}
    \caption{Four main crystal habits in ice clouds (from Ref. \cite{Yang2005})}\label{habits}
\end{figure}

Even though ice clouds are composed of a myriad of crystal habits, we can group them in four main shapes  showing hexagonal prismatic faces due to the hydrogen molecular bonds \cite{Yang2005}, namely droxtal (shape used to represent the complexity of smaller particles with  not perfectly spherical surface), plates, columns and  rosettes as pictured in Fig. \ref{habits}. Then there are intermediate shapes, such as bullets and cups. These shapes grow through deposition/sublimation when turbulent motions are not intense; otherwise, aggregation and collisional mechanisms arise, leading to the formation of more asymmetric shapes, which are generally referred to as aggregates. However, all crystals start growing at the beginning of the nucleation process in a deposition ``regime'' from a condensation nucleus (CN), which can be a small particle of a pollutant or a water droplet, with dimensions ranging from a few hundred nanometres to a few microns. The growth process can be initiated either when water vapour condenses and freezes onto the CN, or when an already frozen  crystal comes into contact with a small water droplet, causing it to freeze. Even though the smaller the sizes  the more  the shape tends to be spherical, the crystal does not assume a perfectly smooth surface because of the water hydrogen bonds, and can be modeled as droxtal shape. However, such small ice crystals have also been found  to occur permanently in high concentrations in  the upper portion of cirrus clouds \cite{Yang2003}, where temperature and humidity fall drastically.
Despite everything we have mentioned, as far as we know, there is still no theoretical model or framework that can predict the evolution of ice crystal shape and habit transition in the deposition regime, while also ensuring coherence with the Fokker–Planck (FP) equations that describe size growth. Even though some important studies, such as \cite{Libbrecht2004,Libbrecth2001,libbrecht2020}, have been carried out in the past years shedding light on the processes that regulate the growth of different crystal habits, a complete and adequate framework able to describe the evolution of probability fields associate with the different habits and their mutual interaction is still missing. 

Ice clouds are characterised by significant internal motions and structures at both the mesoscale and the microscale \cite{Sassen1989}. In particular, vertical air motion variability plays a fundamental role in the internal dynamics of the cloud \cite{Yang2013}. Nevertheless, the impact of small-scale supersaturation variability on the microphysical properties is still partially understood. We know that temperature variability, brought about by gravity waves and turbulence leading to rapid buoyancy oscillations, and water vapour pressure variability generated by localised events cause fluctuations in supersaturation inside ice clouds. These stochastic fluctuations represent both the trigger and the driver for the microphysical time evolution. Currently, there is only limited experimental evidence that there are temperature ranges in which crystals tend to have a dominant shape (columnar or plate-like) \cite{Bailey2009,Nakaya1954}. 

Significant advancements in physics have often been achieved through the interchange of theoretical frameworks across different energy scales. For example, techniques developed in high energy physics such as gauge symmetry and group theory have found notable applications in condensed matter physics and statistical physics \cite{anderson1963plasmons, nambu1961dynamical}. Similarly, effective models originating from low energy physics have contributed to a deeper understanding of emergent phenomena in quantum field theory \cite{wen2004quantum}.
The idea is to explore the possibility of applying concepts from particle physics to transport theory, including mechanisms such as symmetry breaking and gauge symmetries \cite{weinberg1979phenomenological, poniatowski2019superconductivity}. In this context, the Fokker–Planck (FP) equation \cite{Fokker1914,Plank1917,Dekker1979,risken1996fokker}, which describes the evolution of probability distributions in stochastic systems, provides an intriguing intersection point for these ideas. Consequently, the growth of different shapes of ice crystals represents an innovative and promising application of the FP equation, where symmetry principles and field-theoretical methods could offer new insights into the microscopic dynamics of crystal habit.

In this work, we explore the concept of gauge symmetry in order to describe the interaction properties of the ice crystal habits, and we propose a heuristic extension of the FP equation. This allows us to naturally embed the habit structure and the transition from one shape to another.

We begin by describing the crystal habit dynamics in terms of gauge symmetries, starting from a decoupled framework with independent U(1) symmetries assigned to each habit. This initial construction allows a simple, yet powerful representation of each crystal habit evolving independently. We then extend this structure by introducing a minimal interacting model based on an SU(2) $\otimes$ U(1) gauge group. This extension captures nontrivial couplings between three of the four main habits, while keeping the fourth dynamically decoupled. The resulting framework modifies the classical Fokker–Planck dynamics by incorporating interaction terms governed by a habit coupling constant, thus allowing for richer phenomenology and interplay among crystal growth modes.

Our approach seeks to explore the potential of this conceptual framework and lays the foundation for a more comprehensive theoretical formulation of FP-like equations closely connected with gauge theories. At this stage, our study focuses exclusively on ice clouds, deliberately excluding the presence and influence of supercooled water droplets as found in mixed-phase clouds; hence, processes such as riming and  Wegener–Bergeron–Findeisen (WBF) are not considered. Furthermore, we concentrate on ice crystal growth in the deposition regime, whereby crystals grow primarily through water vapor deposition and sublimation on their surfaces, driven by supersaturation.

More specifically, the work is organized as follows. In Section \ref{sec2}, we present the general formalism enabling the reconstruction of the Fokker–Planck equation from a gauge-theoretic perspective, considering both Abelian and non-Abelian gauge groups, providing a foundation for mapping stochastic dynamics into a gauge-invariant framework. In Section \ref{sec3}, this construction is applied to the physical system involving four primary ice crystal habits: droxtals, plates, columns, and rosettes. Section \ref{sec4} introduces the physical variables describing the coupled system of equations encoding both growth process and interactions among the crystal habits, representing the core of our study. Finally, in Section \ref{sec5}, we summarize our findings and discuss future directions, including possible extensions to more complex symmetry structures, as the introduction of a mixing term between SU(2) and U(1) groups, the study and numerical solution of the modified equations and the study of the constant coupling $g_H$.

\section{Gauge Theory Approach to Fokker–Planck Equations}\label{sec2}

The general form of the Fokker–Planck equation for a probability density distribution  $P(\textbf{x},t)$ as a function of time $t$ associated with ${\rm M }$ stochastic variables $x_1,\dots,x_M$ represented by the ${\rm M }$-dimensional state vector $\textbf{x}$, is given by \cite{Fokker1914,Plank1917,Dekker1979}:
\begin{equation}\label{FP}
    \frac{\partial P(\textbf{x},t)}{\partial t} = - \sum_i^M \partial_i (D_i(\textbf{x}) P(\textbf{x},t)) + \sum_{ij}^M\mathbb{D}_{ij}\partial_i\partial_jP(\textbf{x},t)\,\,,
\end{equation}
where  $P(\textbf{x},t)$ describes the probability of finding the stochastic variables in the interval [\textbf{x}, \textbf{x}+d\textbf{x}] at time $t$, while the vector $D_i(\textbf{x})$ represents the drift term and $\mathbb{D}_{ij}$ the diffusion tensor, which is assumed not dependent on the variables $\textbf{x}$ as in \cite{Tome2015}. These terms represent the deterministic and stochastic components, respectively.

The FP equation in \eqref{FP} describes the macroscopic evolution of the system  in terms of the $x_i$ variables and can be derived via Kramers-Moyal expansion \cite{Risken1996} or Ito/Stratonovich theories \cite{Gardiner2009} from the set of Langevin equations that govern the microscopic stochastic dynamics of the $x_i$ and are given in the generic form:
\begin{eqnarray}\label{Lang}
       \frac{dx_i}{dt}= D_i(x_1,x_2,...,x_M) + \xi_i(t)\,\,,
\end{eqnarray}
where the stochastic variables $\xi_1(t),\xi_2(t),...,\xi_N(t)$ describe the gaussian-markovian (white) noise and have the properties:
\begin{equation}\label{langevin}
     \left\langle\xi_i(t)\right\rangle = 0\,\,\,\,\,\,\,,\,\,\,\,\,\,\,
     \left\langle\xi_i(t),\xi_j(t^{\prime})\right\rangle =2\mathbb{D}_{ij}\delta(t-t')\,\,,
\end{equation}
such that the parameter contained in the diffusion tensor $\mathbb{D}$ represents the amplitude of the noise and $\delta(t-t')$ denotes the Dirac delta function.

As we will show in the following sections, the formulation of the Fokker–Planck equation in Eq.~\eqref{FP} can be mapped within the quantum field theory framework by employing the simple Abelian symmetry group U(1), following an alternative approach of what is detailed in \cite{Montigny2003}. We then explore a generalization to both the Abelian U(1)$^\text{N}$ and the non-Abelian SU(N) case. This leads to a generalized set of coupled Fokker–Planck equations describing higher-dimensional probability structures, which naturally represent the probability distribution of several interacting habits.

\subsection{Abelian gauge symmetry}\label{U1}
\noindent
The first step is to construct a gauge theory invariant under the U(1) group within the quantum field theory framework, such that it maps into the Fokker–Planck equation under appropriate definitions. To achieve this, we can consider a pure Yang–Mills (YM) Lagrangian that is invariant under the Abelian U(1) group, as in the well-known case of electromagnetism:
\begin{equation}\label{YM_Lagrangian}   
\mathcal{L}=-\frac{1}{4} F_{\mu \nu} F^{\mu \nu}\,\,,
\end{equation}
where the tensor $F_{\mu \nu}$ is the field strength tensor $F_{\mu \nu} = \partial_\mu A_\nu - \partial_\nu A_\mu$ in terms of the gauge field $A_\mu$. 
The Lagrangian is built to be invariant on a gauge transformation in the usual form:
\begin{equation}\label{AtransfU1}
    A_\mu \rightarrow A'_\mu = A_\mu - \frac{1}{g_1}\partial_\mu \Lambda\,\,,
\end{equation}
with $\Lambda$  an arbitrary function and $g_1$ is the coupling constant. \\
From the Lagrangian, it is simple to derive the Euler-Lagrange equations that assume the general form:
\begin{equation}
\partial_\mu \partial^\mu A^\nu - \partial_\mu\partial^\nu A^\mu = 0\,\,.
\end{equation}
From the gauge invariance, it is possible to implement a gauge-fixing term as $\partial_\mu \partial^\mu A^\nu=0$. In such a way the equation of the motion is simplified to:
\begin{equation}\label{divA}
 \partial_\mu A^\mu  = 0\,\,.    
\end{equation}
This condition implies the additional constraint on the $\Lambda(\textbf{x})$ function:
\begin{equation}\label{eq:lambdacond}
\partial^{\mu}\partial_{\mu}\Lambda = 0\,\,.
\end{equation}
Notice that the Eqs.~\eqref{divA} and \eqref{eq:lambdacond} represent the opposite choice of what is normally done in Quantum Field Theory in order to proceed to the quantization of the photon field. However Eqs.~\eqref{divA} and \eqref{eq:lambdacond} represent a convenient choice for deriving the FP equation and they are commonly used in the literature (see for instance Ref. \cite{Montigny2003} and references therein). It is worth noticing that our theory is a classical theory of fields (the so called second quantization is not employed here).
Now, in order to parametrize the field $A_\mu$ \cite{Montigny2003} to reconstruct the FP equation we request the following definitions:
\begin{equation}\label{fieldA_FP}
\begin{aligned}
    A^i(\textbf{x},t) &=  D^{{i}}(\textbf{x}) P(\textbf{x},t)  -  \mathbb{D}^{ij}\partial_j   P(\textbf{x},t) \,\,,\\
    A^0(\textbf{x},t) &= P(\textbf{x},t)\,\,,    
\end{aligned}    
\end{equation}
where $D_i(\textbf{x})$, $\mathbb{D}_{ij}$ and $P(\textbf{x},t)$ are respectively the drift, the diffusion tensor and the density probability as introduced in  Eq. \eqref{FP}. For simplicity, we will refer to $A_\mu$ as the Fokker–Planck field (or FP field).

Using  Eq. \eqref{fieldA_FP} and the contravariant derivative $\partial_\mu= (\partial_t,\partial_i)$, we can explicitly obtain the FP equation equivalent to Eq. \eqref{FP}:
\begin{equation}
\label{eq:Fokker-Plank}
    \partial_t P(\textbf{x},t) +\partial_i (D_i(\textbf{x})P(\textbf{x},t))  - \mathbb{D}_{ij}\partial_i \partial_jP(\textbf{x},t) =0\,\,,
\end{equation}
where we considered the covariant definition for the field $A^\mu = (A^0,A^i)$ and the metric tensor is implicitly used, $\mathbb{g}_{\mu\nu}=\text{diag}(1,1,\ldots,1)$, where the ``$\ldots$'' manifests the possibility to have in addition to the time $t$, $M$ possible extra stochastic variables, that could not be related to space-time dimensions (see Sec.~\ref{sec4} for an explicit example). So we have shown that it is possible to obtain the FP equation starting from a U(1) gauge theory. Note that, compared to Eq.~\eqref{FP}, we have omitted the summation symbol in Eq.~\eqref{eq:Fokker-Plank}. Repeated indices imply summation over them.
A couple of comments are in order regarding the procedure explained in Ref.~\cite{Montigny2003}. That work arrives at the same FP equation as Eq.~\eqref{eq:Fokker-Plank}, but through a different approach: whereas our metric tensor is flat, the stochastic metric tensor defined in Eq.~(17) of Ref.~\cite{Montigny2003} depends on the coordinates and extends the geometric structure into a five-dimensional manifold. This has a profound impact on the definition of the gauge fields $A^\mu$. While in that work the fields are independent of the diffusion tensor $\mathbb{D}_{ij}$ (its presence in the FP equation is recovered via the coordinate-dependent metric), in our case Eq.~\eqref{fieldA_FP} provides the full dependence on both the drift term and the diffusion tensor. The framework in Ref.~\cite{Montigny2003} embeds stochastic dynamics within the geometric structure itself and our approach preserves a classical, fixed spacetime geometry. This allows for a mapping between relativistic field-theoretic and stochastic formulations rather than unifying or geometrically embedding the two frameworks.

The FP equation \eqref{eq:Fokker-Plank} can be recast in a probability conservation expression. This can be written as a continuity equation introducing a probability current:
\begin{equation}
    J_i(\textbf{x},t) = D_i(\textbf{x})P(\textbf{x},t)  -  \mathbb{D}_{ij}\partial_j P(\textbf{x},t) \,\,,
\end{equation}
which assumes the form
\begin{equation}
     \partial_t P(\textbf{x},t) = -\partial_i J_i (\textbf{x},t)\,\,.
\end{equation}
The total probability conservation therefore requires: 
\begin{equation} \label{cons_prob}
     \frac{d}{dt}\int_{\Omega} P(\mathbf{x}, t) \, d\mathbf{x} = 0\,\,.
\end{equation}
Eq. \ref{cons_prob} expresses a fundamental property because it ensures that the probability distribution remains normalised at any time. This condition is fulfilled under suitable boundary conditions \cite{Tome2015}. For example, if $ \Omega $ is a finite domain, the integral of the divergence of the probability current (the integral in the left hand side of Eq.~\eqref{cons_prob}) must vanish at the boundaries.


For our purposes, we now extend the Fokker--Planck equation to ${\rm N }$ types of crystal habits. The simplest way to achieve this is by introducing an enlarged internal symmetry group U(1)$^\text{N}$, where each U(1) corresponds to an independent gauge symmetry associated with a specific crystal habit.

Starting from this symmetry structure, we can generalize the Lagrangian in Eq.~(\ref{YM_Lagrangian}) to include ${\rm N }$ different gauge fields as follows:
\begin{equation}
    \mathcal{L}= -\frac{1}{4} \sum_{h=1}^N  F^h_{\mu \nu} F_h^{\mu \nu} \qquad\,\,, \qquad h=1, ..., {\rm N }\,\,,
\end{equation}
with a field strength tensor $F^h_{\mu \nu}$ associated to the $h$-th habit. Following the same approach as in the single-habit case, this leads to ${\rm N }$ separate equations of motion, and hence ${\rm N }$ FP-like equations, one for each habit, completely decoupled:
\begin{equation}
\label{eq:U1allaNmatrix}
\left\{
\begin{aligned}
    &\partial_t P_1(\textbf{x},t) +\partial_i \left(D^{(1)}_i(\textbf{x}) P_1(\textbf{x},t)\right) - \mathbb{D}^{(1)}_{ij}\partial_i \partial_jP_1(\textbf{x},t)  = 0\,\,, \\
    &\partial_t P_2(\textbf{x},t)  + \partial_i \left(D^{(2)}_i(\textbf{x}) P_2(\textbf{x},t)\right) - \mathbb{D}^{(2)}_{ij}\partial_i \partial_jP_2(\textbf{x},t) = 0\,\,, \\
    &\hspace{50pt} \vdots \\
    &\partial_t P_N(\textbf{x},t)  + \partial_i \left(D^{(N)}_i(\textbf{x}) P_N(\textbf{x},t)\right) - \mathbb{D}^{(N)}_{ij}\partial_i \partial_j P_N(\textbf{x},t)= 0\,\,. \\
\end{aligned}
\right.
\end{equation}
In analogy with the single-habit case, we introduce ${\rm N }$ probability currents:
\begin{equation}
    J^h_i(\textbf{x},t) = D^{(h)}_i(\textbf{x})P^h(\textbf{x},t)  -  \mathbb{D}^{(h)}_{ij}\partial_jP^h(\textbf{x},t)\,\,, 
\end{equation}
which satisfy the continuity equations
\begin{equation}
    \partial_t P^h(\textbf{x},t) = \partial_i J^h_i(\textbf{x},t) \qquad ;\quad \forall\, h = 1, \dots, {\rm N }\,\,.
\end{equation}
We can then define a total probability function as the average over the individual probabilities:
\begin{equation}
    \mathcal{P}(\textbf{x},t) = \frac{1}{N}\sum_{h=1}^N P_h(\textbf{x},t)\,\,.
\end{equation}
To ensure conservation of the total probability, we impose the condition:
\begin{equation}
     \frac{d}{dt} \int_{\Omega} \mathcal{P}(\textbf{x},t)\, d\textbf{x} = \frac{1}{N} \sum_{h=1}^N \frac{d}{dt} \int_{\Omega} P_h(\textbf{x},t)\, d\textbf{x} = 0\,\,,
\end{equation}
which is satisfied provided that each individual probability $P_h$ is conserved (since the N habits are completely decoupled), as discussed in the single-habit case.

\subsection{Non-Abelian gauge symmetry}\label{Nonabel}
\noindent 
In this section, we aim to introduce appropriate mixing between the habits. To achieve this, we will extend the symmetry to the non-Abelian case (specifically the SU(N) group) following \cite{Montigny2003}\footnote{Notice that there are substantial differences between the approach followed in Ref. \cite{Montigny2003} and what is performed in this work. See discussion below Eq.~\eqref{eq:Fokker-Plank}.} . We proceed starting from the YM Lagrangian that is invariant under a local gauge transformation of this group,
\begin{equation}
    \mathcal{L}= -\frac{1}{4}  F_{\mu \nu}^{a} F^{\mu \nu}_{a}\,\,,
\end{equation}
where $a$ is the index which runs over the group generators ($a=1, 2, \ldots, {\rm N }^2-1$), and the summation over repeated internal indices of SU(N) is always understood. In this case the strength tensor $F^a_{\mu \nu}$ is written in terms of the three gauge FP fields $A_\mu=A_\mu^a T^a$, where $T^a$ are the SU(N) group generators (in the case in which N $=2$, for the fundamental representation of SU(2) the generators $T^a$ are related to the well known Pauli matrices $\sigma^a$ thorough $T^a=\sigma^a/2$, with $a=,1,2,3$). The strength tensor in terms of $A_\mu^a$ assumes the form:
\begin{equation}
\label{eq:FmunuNA}
    F^{a}_{\mu \nu} = \partial_\mu A^a_\nu - \partial_\nu A^a_\mu+ i g[A_\mu,A_\nu]^a
    =\partial_\mu A^a_\nu - \partial_\nu A^a_{\mu}-  g f^a_{bc}A^b_{\mu} A^c_{\nu} \,,
\end{equation}
where $f_{abc}$ are the completely antisymmetric structure constants of SU(N). In the particular case of N $=2$, $f_{abc} = \epsilon_{abc}$ are the totally antisymmetric Levi-Civita tensor (see Appendix \ref{sec:Gaugegroups}). The generators of SU(N) $T_a$ satisfy the Lie algebra commutation relations:
\begin{equation}
   [T_a, T_b] = i f_{abc} T_c, \qquad f_{abc} = -f_{bac}\,\,.
\end{equation}
In the non-Abelian case, the Lagrangian is built to be invariant on the following gauge transformation:
\begin{equation}\label{AtransfSU2}
A_\mu \ \rightarrow A'_\mu = \ U A_\mu U^\dagger  + \frac{i}{g}  U^\dagger (\partial_\mu U)
\end{equation}
where $U(x) = e^{i\Lambda^a(\textbf{x}) T^a} $ is a local transformation under the SU(N) group (see Appendix \ref{sec:Gaugegroups} for more details), $\Lambda^a$ are arbitrary well-behaved functions of the space-time coordinates and $g$ is the coupling of the field theory. Taking  Eq.~\eqref{AtransfSU2} and considering an infinitesimal transformation, we have:
\begin{equation}\label{AtransfSU2}
A^a_\mu \ \rightarrow  A^{\prime a}_\mu= A^a_\mu -\frac{1}{g} \partial_{\mu}\Lambda^a -  f_{abc}\Lambda^bA^c_{\mu}\,\,.
\end{equation}
From the precedent equation and imposing the gauge-fixing condition $\partial_\mu\partial^\mu A_\nu^a = 0$ we arrive to a constraint on the $\Lambda_a$ functions,
\begin{equation}\label{gauge2}
\partial_\mu \partial^\mu \Lambda^a = - g \int f^{abc} \partial_\mu \partial^\mu \left( \Lambda^b A_\nu^c \right) dx^\nu\,.
\end{equation}
The Euler-Lagrange equations in this non-Abelian case have the form:
\begin{equation}
\label{eq:cov_der}
    \Delta_{\mu} F^{\mu \nu} = 0\,,
\end{equation}
where we have introduced the following covariant derivative,
\begin{equation}
(\Delta_\mu )_{ab} = \partial_\mu \delta_{ab} + g f_{abc} A^c_\mu \,. 
\end{equation}
Inserting Eq.~\eqref{eq:FmunuNA} in Eq.~\eqref{eq:cov_der} we can explicitly derive the equations of motion:

\begin{equation}
\begin{aligned} 
\label{eq:E-L-NA}
     \partial^\nu\partial_\mu A^\mu_a = g\,f_{abc}\,\left(-\partial_\mu(A^\mu_b A^\nu_c) \,+\, F^{\mu\nu}_b\,A_\mu^c\right)\,.
\end{aligned}   
\end{equation} 
where we have already used the gauge fixing condition $\partial_\mu\partial^\mu A_\nu^a = 0$. Notice that the second term in the right-hand side of Eq.~\eqref{eq:E-L-NA} coincides with the SU(N) current associated with the conserved SU(N) charge (see Appendix \ref{AppendixC}). As it is usual in Quantum Field Theory, the associated current works as a \textit{source} for the equations of motion of the corresponding field. This is the direct interaction of the strength of one field on the potential of another. The first term in the right-hand side of Eq.~\eqref{eq:E-L-NA} can be understood as a source arising at the boundary between regions of high and low interaction. This term is sensitive to the field's behavior in a neighborhood (due to the presence of the four-divergence). The derivative $\partial_\mu$ measures how this interaction potential changes in spacetime. In terms of the physics of crystal habits, consider a region within a cloud where the physical conditions, such as temperature and supersaturation, vary. In one part of this region, the conditions are favourable for the interaction between the field of plates (habit ``b'') and the field of columns (habit ``c''). Here, the ``interaction potential'' $A_b A_c$ is high. In an adjacent area, however, the conditions are different, and this interaction is much weaker. The potential $A_b A_c$ is low. The first term tells us that the boundary or gradient between these two parts (the place where the interaction strength changes) acts as a source for forming, for instance, rosettes (habit ``a''). It is not just the presence of the other habits, but the non-uniformity of their interaction that drives the formation of the third. 

Eq.~\eqref{eq:E-L-NA} can be recast in the following form
\begin{equation}
\begin{aligned} 
\label{eq:FP-NA}
     \partial_\mu A^\mu_a =f_{abc} \left(g\,\, \mathcal{I}^{(1)}(\mathbb{A}^b,\mathbb{A}^c) + g^2\,\, \mathcal{I}^{(2)}(\mathbb{A}^b,\mathbb{A}^c) \right)\,.
\end{aligned}   
\end{equation}
The first term in the left-hand side of Eq.~\eqref{eq:FP-NA} resembles the FP equation in the U(1) Abelian case for each internal SU(N) index $a$ as already deduced for the Abelian case in Eq.~\eqref{divA}. Therefore the right-hand side of Eq.~\eqref{eq:FP-NA} can be regarded as ``the \textit{mixing} terms'' between the habits. For vanishing couplings $g=0$ it is possible to recover the Abelian case for each one of the fields $A^\mu_a$ being completely decoupled (without mixing). Notice that for the choice of the group representation made in this section, we have always N$^2-1$ fields or habits \footnote{The fact that only N$^2-1$ fields or habits can be considered for each case could be a limiting aspect of our approach. However after fixing the dimensionality of the group ${\rm N }$ it is possible to reduce or even enlarge the number of habits as a matter of convenience (see next section). }. In the left-hand side of Eq.~\eqref{eq:FP-NA} the integral operators $\mathcal{I}$ are written in term of the $A^\mu_a$ fields and possibly (at most) first order derivatives (this fact is denoted with the typographical convention $\mathbb{A}^a$). The explicit order of the coupling accompanying each of these integral terms $\mathcal{I}^{(i)}$ is denoted with the superscript $i=1,2$. Therefore $\mathcal{I}^{(1)}$ collects all the $\mathcal{O}(g)$ terms from the right hand side of Eq.~\eqref{eq:E-L-NA} and therefore the corresponding second order terms $\mathcal{O}(g^2)$ are collected in $\mathcal{I}^{(2)}$. For the sake of generality (without specifying the size of the coupling $g$ and/or the number of habits, etc.) we prefer to do not make any assumption in the symmetry of the solutions of the non-Abelian FP equation \cite{Montigny2003} or the stationary condition of the fields to further simplify right-hand side of Eq.~\eqref{eq:FP-NA}. Taking the assumptions made in Ref.~\cite{Montigny2003} our Eq.~\eqref{eq:FP-NA} reduces to Eqs.~(23) and (24) of that work.

Explicitly the $\mathcal{I}^{(i)}$ integral operators read
\begin{align}
    \mathcal{I}^{(1)}(\mathbb{A}^b,\mathbb{A}^c) & = \int {\rm d}x^\nu  \left[A_\mu^c\,\left( \partial^\nu A^\mu_b - \partial^\mu A^\nu_b \right) - \partial_\mu\left( A^\mu_b A^\nu_c\right)\right] \,\,,\\
    \mathcal{I}^{(2)}(\mathbb{A}^b,\mathbb{A}^c) & = - \int {\rm d}x^\nu  f_{blm}\, A^\mu_l \, A^\nu_m \, A_\mu^c \,\,.
\end{align}
In order to evaluate the impact of the SU(N) group in the modified FP equation, we generalize the definition of the FP gauge field $A_{a}^{\mu}$ from the one written in Eq. (\ref{fieldA_FP}) for each internal SU(N) index $a$:
\begin{equation}
\begin{aligned}\label{Amu}
    A_a^i(\textbf{x},t) &=  D^i_{{(a)}}(\textbf{x}) P_a(\textbf{x},t)   -  \mathbb{D}_{(a)}^{ij} \partial_j   P_{a}(\textbf{x},t)\,\,,  \\
    A^0_a(\textbf{x},t) &= P_{a}(\textbf{x},t)\,.    
\end{aligned}    
\end{equation}
When the expression for the $A^{\mu}_a$ fields, given in Eq.~\eqref{Amu}, is inserted into Eq.~\eqref{eq:FP-NA}, the latter reduces to the standard FP equation (for each internal SU(N) index), plus a new contribution proportional to the coupling constant $g$ (terms up to order $g^2$ are present only). These terms represent the typical consequence of the non-Abelian nature of the theory (they are absent in the Abelian case).

For the sake of clarity, we omit the explicit expression of the integral operators $\mathcal{I}^{(i)}$, which can be found in Appendix \ref{AppendixB}. Here we present the FP equation containing the new terms inherited from the SU(N) nature of the fields in the following form:
\begin{equation}\label{FPnnn}
\partial_t P_a(\mathbf{x},t) + \partial_i\bigl(D_{(a)}^{i}(\mathbf{x})P_a(\mathbf{x},t)\bigr) - \mathbb{D}_{(a)}^{ij}\partial_i \partial_j P_a(\mathbf{x},t) = f_{abc} \bigl(g\,\mathcal{I}^{(1)}(\mathbb{A}^b,\mathbb{A}^c) + g^2\,\mathcal{I}^{(2)}(\mathbb{A}^b,\mathbb{A}^c)\bigr)\,,
\end{equation}
where the internal SU(2) index $a=1,2,\ldots,N^2-1$. These additional terms demonstrate how the probability of a particular habit may change due to the influence of others. This can be understood by considering the mix between the internal SU(N) indices. As discussed in Section \ref{U1} regarding the U$^{\rm N}$(1) theory, there is no mixing between the internal U(1)$^{\rm N}$ indices in any of the N equations in expression \eqref{eq:U1allaNmatrix}. The left-hand side of Eq.~\eqref{FPnnn} resembles the left-hand side of any of the equations in Eq.~\eqref{eq:U1allaNmatrix} for each internal SU(N) index ``\textit{a}''. Meanwhile, in the right-hand side of Eq.~\eqref{FPnnn} the presence of the integral operators $\mathcal{I}^{(i)}$ (convoluted with the structure constants of SU(N)) on the right-hand side of Eq. ~\eqref{FPnnn} provides the ``\textit{interaction}'' terms that mix the internal SU(N) indices. This new expression can be understood as interaction terms between different habits or as a habit changing shape and evolving into a new type.

\section{Crystals habit gauge symmetry}\label{sec3}
In this section, we extend the Fokker–Planck equation in order to describe the dynamics of four distinct types of crystal habit: droxtals, plates, columns and rosettes. To capture the mutual influence among the different habits, without assigning any preferential role, we adopt an extended gauge group framework that reflects the underlying symmetry of the system.

The simplest scenario corresponds to a configuration in which the crystal habits evolve independently, with no interactions between them. This is naturally described by an Abelian product symmetry represented by the group U(1)$^4$, where each habit is assigned its own independent U(1) gauge symmetry (see Section \ref{U1}). The structure of this symmetry group can be schematically expressed as:
\begin{equation}
    U(1)^4: \left(\begin{array}{cccc}
     U(1)_{\text{column}}&  & & \\
     & U(1)_{\text{plate}}& & \\
     &  & U(1)_{\text{bullet rosette}} & \\
     &   &  & U(1)_{\text{droxtal}}\\
\end{array}\right)\,\,,
\end{equation}
which provides an abstract representation of independent symmetries for each habit. In this case, the Lagrangian takes the form:
\begin{equation}
    \mathcal{L}= -\frac{1}{4} \sum_{h=1}^4  F^h_{\mu \nu} F_h^{\mu \nu}\,\,, \qquad\,\, \qquad h=1, ...,4\,\,,
\end{equation}
where each $F^h_{\mu \nu}$ is the field strength tensor associated with the $h$-th habit. From this Lagrangian, using the usual approach, we obtain four distinct FP equations, one for each type of habit:
\begin{equation}
\partial_t P_h(\textbf{x},t) + \partial_i \left(D^{(h)}_i(\textbf{x}) P_h(\textbf{x},t)\right) - \mathbb{D}^{(h)}_{ij}\partial_i \partial_j P_h(\textbf{x},t) = 0\,\,, \qquad \text{for } h = 1, ..., 4\,\,.
\end{equation}
However, if we wish to introduce interactions between the different types of habit, a bigger symmetry group could be considered (as it was studied in Section \ref{Nonabel}). One possible minimal choice is given by the group SU(2) $\otimes$ U(1), that allows to embed four habits with a dynamic coupling among three habits leaving the fourth one decoupled. The structure of this group can be schematically represented as:
\begin{equation}
    SU(2)\otimes U(1): \left(\begin{array}{ccc|c}
        &  & & \\
     & SU(2)& & \\
     &  & & \\
    \hline
     &   &  & U(1)\\
\end{array}\right)\,.
\end{equation}
In this configuration, the first three habits are coupled through the SU(2) group, while the fourth remains isolated, governed only by a U(1) symmetry. The corresponding Lagrangian takes the form:
\begin{equation}
\mathcal{L}= -\frac{1}{4}\mathcal{G}^{h}_{\mu\nu}\mathcal{G}_{h}^{\mu\nu} \equiv -\frac{1}{4} F_{\mu \nu}^{a} F^{\mu \nu}_{a} -\frac{1}{4} W_{\mu \nu} W^{\mu \nu}\,, \qquad\qquad a=1,\cdots,3\,;\hspace{0.2cm} \quad h=1,\cdots, 4 \,,
\end{equation}
where $\mathcal{G}_{h}^{\mu\nu} \equiv (F_a^{\mu\nu},W^{\mu\nu})$, and the field strength tensors are defined as:
\begin{equation}
F^{a}_{\mu \nu} = \partial_\mu A^a_\nu - \partial_\nu A^a_{\mu}-  g_H \epsilon^a_{bc}A^b_{\mu} A^c_{\nu}\,\,, \qquad W_{\mu \nu} = \partial_\mu B_\nu - \partial_\nu B_\mu\,,
\end{equation}
and $g_H$ is the habit coupling constant of SU(2).

The equations of motion derived from the Lagrangian become:
\begin{equation}
    (\Delta_{\mu})_{hh'}\mathcal{G}^{\mu\nu}_{h'} = 0\,,
\end{equation}
were the covariant derivative is defined as:
\begin{equation}
    (\Delta_{\mu})_{hh'} = \left(\begin{array}{ccc|c}
     \partial_{\mu}   & g_HA^3_{\mu} & -g_HA^2_{\mu} & 0\\
     -g_HA^3_{\mu} & \partial_{\mu}&g_HA^1_{\mu} & 0\\
    g_HA^2_{\mu} & -g_HA^1_{\mu} & \partial_{\mu}& 0\\
    \hline
    0 &  0 &0  & \partial_{\mu} \\
\end{array}\right)\,, \qquad \qquad\forall h,h'=1,\cdots,4    \,.
\end{equation}
This structure leads to a modified FP system:
\begin{equation}\label{eqmotion}
\begin{cases}
\begin{aligned}
    \partial_t P_a(\textbf{x},t) &= -\partial_{i}D^{(a)}_{i} P_{a}(\textbf{x},t) + \mathbb{D}^{(a)}_{ij}\partial_i\partial_jP_a(\textbf{x},t) + f_{abc} \left(g_H\,\, \mathcal{I}^{(1)}(\mathbb{A}^b,\mathbb{A}^c) + g_H^2\,\, \mathcal{I}^{(2)}(\mathbb{A}^b,\mathbb{A}^c) \right)\,,\\
    \partial_t P_4(\textbf{x},t) &= -\partial_iD^{(4)}_{ i} P_{4}(\textbf{x},t) + \mathbb{D}^{(4)}_{ij}\partial_i\partial_jP_{4}(\textbf{x},t)\,. \\
\end{aligned}
\end{cases}
\end{equation}
In this system, the first three habits (see Eq.~\eqref{FPnnn}) interact with each other through the $g_H$ coupling terms, while the fourth evolves independently (see Eq.~\eqref{eq:U1allaNmatrix} and discussion below Eq.~\eqref{FPnnn}). The introduction of interactions requires a conservation condition on the total probability:
\begin{align}
     \frac{d}{dt}\sum_{a=1}^3 \int_{\Omega} P_a(\textbf{x},t) \, d\textbf{x}  = 0\,\,, \qquad \qquad 
     \frac{d}{dt} \int_{\Omega} P_4(\textbf{x},t) \, d\textbf{x}  = 0 \,\,.
\end{align}
We have seen that enlarging the gauge group leads to the appearance of additional linear and quadratic terms in the coupling constant $g_H$ within the Fokker–Planck equation (see section \ref{Nonabel}). These terms introduce contributions that encode interaction effects among different crystal habits. In the limit $g_H \to 0$, where such interactions vanish, the system consistently reduces to the Abelian U(1)$^4$ case.

In our minimal interacting model, three of the four crystal habits, plates, columns and bullet rosettes, are associated with the SU(2) sector of the gauge group, while the droxtal habit is linked to the U(1) component. This assignment reflects the idea (or assumption) that the droxtal generally represents an early stage of crystal formation, typically involving uniform growth without a privileged dimension \cite{Gonda1978}. This habit can also represent particles at very low temperatures and humidity in the higher part of cirrus clouds.
Within the SU(2) $\otimes$ U(1) structure, we assume no explicit mixing between the two symmetry sectors. As a result, the droxtal evolves independently of the other habits. This corresponds to a simplified but physically motivated assumption, where the subspaces associated with SU(2) and U(1) remain invariant and dynamically decoupled.

A possible extension of this framework could involve the introduction of interaction terms in the Lagrangian, proportional to expressions like $A_\mu^a B^\mu$. These terms would couple the field associated with the droxtal to those of the other three habits. Such coupling would enable the transition from the small, few-micron crystal state (the droxtal, associated with the U(1) subspace) to more evolved stages of growth \cite{Libbrecht2005}. The interaction would allow the system to access higher-dimensional and larger-size configurations, corresponding to plates, columns and rosettes, which are embedded in the SU(2) sector of the gauge group. Such terms would explicitly break the SU(2) $\otimes$ U(1) symmetry, which would then survive only as a residual symmetry under specific conditions. However, this possibility lies beyond the scope of the present analysis.

\section{Introducing physical variables $\alpha$ and $\varphi$ describing the habit growth} \label{sec4}
The dynamic of an ice crystal  growth, that represent the depositional regime,  is completely determined  by  the supersaturation $s$, which is defined by:
\begin{equation}\label{s}
    s(T) = \frac{p_v}{p_s(T)}-1 = \frac{q_v}{q_s(T)}-1
\end{equation}
where $q_v$ and $q_s$ are vapor mixing ratio and the saturation vapor mixing ratio, while $p_v$ is the vapor pressure and $p_s$ is the saturation vapor pressure.  The saturation vapor pressure depends on temperature following the empirical law \cite{Marti1993}: 
\begin{equation}
    p_s(T) =  3.4452\cdot10^{10}e^{-Q/T}
\end{equation}
valid within $170 < T < 250$ K, giving $p_s$ in millibars and with $Q=6132.9$ K being the latent heat of sublimation. 
Supersaturation can vary between -1 and, theoretically, $+\infty$; when it is negative we have sublimation and size reduction, and when positive we have deposition and size enhancement. 

The morphology of the crystal is generally not spherical and depends on both the temperature and the supersaturation level, which select specific growth habits (e.g., columns, plates, rosettes). Each habit $h$ can be geometrically characterized by an \textit{equivalent radius} $r_e^{(h)}$, defined as the radius of a sphere with the same volume:
\begin{equation}
 \frac{4}{3}\pi\left(r_e^{(h)}\right)^3 = V^{(h)}\left(l^{(h)}, d^{(h)}\right),   
\end{equation}
where $V^{(h)} $ is the volume of a crystal with a generic habit $h$ , and $l^{(h)}$ and $ d^{(h)} $ denote the length and width of the crystal, respectively~\cite{Karcher2003}.
The velocity of crystal growth, expressed as the time derivative of the effective radius $r_e^{(h)}$, follows the relation \cite{Korolev2003}:
\begin{equation}\label{derre}
    \frac{dr_e^{(h)}}{dt} = \frac{C^{(h)}_{0}A}{r_e^{(h)}} \ s \,,
\end{equation}
where  $C_0^{(h)} $ is the capacitance for the specific habit $h$  and takes into account the non-spherical shape of the crystal. Its values varies between $0<C^{(h)}_{0}\leq1 $ and can be calculated as \cite{Westbrook2006}:
\begin{equation}
C_0^{(h)} = 
\begin{cases}
1 & \text{for spherical}\,, \\
\displaystyle \frac{0.58 \left(1 + 0.95\mathcal{A}_{(h)}^{0.75}\right)}{\mathcal{A}_{(h)}} & \text{for columns and plates} \,,\\
0.8\,\mathcal{A}_{(h)}^{-0.25} & \text{for bullet rosettes}\,,
\end{cases}
\end{equation}
where $\mathcal{A}_{(h)} = d^{(h)}/l^{(h)}$ is the aspect ratio of a single habit  \cite{Solch2010}. The coefficient 
$A=A(T,D_v, L)$ in Eq. \eqref{derre} is defined as \cite{Korolev2003}:\begin{equation}\label{Aterm}
A(T,D_v, L)= \bigg(\frac{\rho L^2}{R_v k T^2}+\frac{\rho R_vT}{p_s(T)D_v(p,T)}\bigg)^{-1}\,,
\end{equation}
where $L$ in the latent heat, $R_v$ the gas constant for moist air, $k$ the heat conductivity, $\rho = 917$ kg/m$^3$ the ice density and $D_v$ the  diffusivity coefficient of water vapour that can be approximated with the following function \cite{Pruppacher1997}:
\begin{equation}\label{Dv}
 D_{v}(p,T)=0.211\bigg(\frac{1013.25}{p}\bigg)\bigg(\frac{T}{273.15}\bigg)^{1.94} \,.
\end{equation}

The evolution of supersaturation can be described as fluctuations around a mean value. Considering only first-order fluctuations, we write $s = \hat{s} + \varphi$ \cite{Karcher2014}. The average supersaturation $\hat{s}$ is given by Eq.~\eqref{s} as a function of the mean local temperature and vapor pressure (or mixing ratio). Rapid fluctuations $\varphi$ around the mean value $\hat{s}$ are driven by perturbations (i.e., noise) in the temperature $T$ and the partial pressure of water vapor $p_v$.  They are damped by the uptake of water molecules on the surfaces of ice crystals and characterized by a finite correlation timescale \cite{Karcher2012}.

As shown in Eq.~\eqref{derre}, the growth velocity of ice crystals depends linearly on supersaturation. The fluctuations $\varphi$ are assumed to evolve on shorter temporal (and spatial) scales than the mean component $\hat{s}$ \cite{Karcher2012}. As noted in the referenced work, the mean supersaturation $\hat{s}$ can be influenced by large-scale background forcings and long-timescale correlation processes, and can be computed by numerically solving the Navier–Stokes equations. Since damping processes within ice clouds generally drive $\hat{s}$ toward thermodynamic equilibrium, it can be treated as an externally provided parameter. We then consider a volume of air consisting of a sufficiently large number of parcels, such that their collective behavior results in a well-defined mean value.
The equation ~\eqref{derre} in the Lagrangian frame ($\hat{s}=0$) can be expressed in terms of the parameters  $\alpha^{(h)} = (r_e^{(h)}/r^{(h)}_{ 0})^2$ which represents the effective crystal surface  scaled by the initial value $r_{0}^{(h)}=r_e^{(h)}(t=0)$ \cite{Karcher2014}:
\begin{equation}\label{eq1_effsurface}
\frac{d\alpha^{(h)}}{dt}  =  \gamma^{(h)} \varphi  \qquad \text{with} \qquad \gamma^{(h)} = \tfrac{2C^{(h)}_{0}A}{r_{0}^{(h)2}}\,.
\end{equation}
This can be coupled with the associated Langevin equation (see Eq.~\eqref{Lang}) for the supersaturation variability $s$ above the surface of the ice crystal in the same frame. 
\begin{eqnarray}\label{eq2_supersat}
\frac{d\varphi^{(h)}}{dt}  =  -\lambda^{(h)} \varphi + \xi(t) \qquad \text{with} \qquad \lambda^{(h)} = \tfrac{4\pi N C_0^{(h)}r_e^{(h)}A  \rho}{\rho_a} \bigg(\tfrac{pR_v}{p_s(T)R_a}+\tfrac{L^2}{c_pR_v T^2}\bigg)\,.
\end{eqnarray}
with  $\rho_a$ and $c_p$ the density of dry air and  and the  specific heat of air at constant pressure  and $\xi(t)$ is the time-dependent random forcing of $\varphi$ that is characterized by a zero mean, exponential correlation over the time $t$ and a Gaussian probability distribution (see Eq. \eqref{langevin}):
\begin{equation}
     \left\langle\xi(t)\right\rangle^{(h)} = 0\,\,\,;\hspace{1cm} 
     \left\langle\xi(t),\xi(t^{\prime})\right\rangle^{(h)} =2D^{(h)}\delta(t-t^{\prime})\,,
\end{equation}
where $D^{(h)}$ is defined as:
\begin{equation}
    D^{(h)}=\frac{D_{v}^{'(h)}}{\frac{D_{v}^{'(h)} L p_s(T)}{k^{'(h)}T}\big(\frac{L}{R_vT}-1\big) +R_vT}\,, \\
\end{equation}
with: 
\begin{align}
    D^{'(h)}_{v} = f^{(h)}c^{(h)}_{h}D_v \\
    k^{'(h)} = f^{(h)}c^{(h)}k
\end{align}
whereby the shape and ventilator coefficients $c^{(h)}$ and $f^{(h)}$. For a single habit $(h)$ the ventilator coefficient is expressed as \cite{Solch2010}:
\begin{equation}
    f^{(h)}=1+p_{1}^{(h)}X_{h}+p_{2}^{(h)} X^{(h)2}\,,
\end{equation}
for plates, columns and bullet rosettes, with $p_{1}^{(h)}$ and $p_{2}^{(h)}$ tabulated in \cite{Solch2010}, and $X^{(h)}$ expressed as:
\begin{equation}
X^{(h)}=S_c^{\frac{1}{3}}(R_{e}^{(h)})^{\frac{1}{2}}=\big(\frac{\eta}{D_v}\big)^{\frac{1}{3}}(a_1Y^{(h)b_1})^{\frac{1}{2}}\,.
\end{equation}
$S_c$ and $R_e$ denote the Schmidt and Reynolds numbers, respectively, $\eta$ is the kinematic viscosity, $a_1$ and $b_1$ are coefficients calculated in  \cite{Solch2010}  and the Best number $Y^{(h)}$ is expressed for each habit as:
\begin{equation}
    Y_{(h)}=\frac{8 g\eta^2}{\rho_a}(r_e^{(h)})^2\frac{m^{(h)}}{A^{(h)}}\,,
\end{equation}
with $g$ the gravity acceleration, $m^{(h)}$ and $A^{(h)}$ indicate the crystal mass and the projected area for the habit $(h)$.  
Finally the shape factors $c^{(h)}$ are given by:
\begin{equation}\label{chh}
    c^{(h)}=\bigg(\frac{r_e^{(h)}}{r_e^{(h)}+\zeta(p,T)}+\frac{2D_{v}C_{0}^{(h)}}{\hat{v}\omega_{v}r_e^{(h)}}\bigg)^{-1}\,,
\end{equation}
where  $\hat{v}=\hat{v}(T)$ denotes the mean thermal speed of water molecules in the vapor, $\omega_{v}$ the deposition coefficients for water vapor uptake on ice crystals which can be set to 0.5 as pointed out in \cite{Solch2010} in agreement with laboratory and field studies \cite{Haag2003},  $D_v=D_v(p,T)$ is the uncorrected gas diffusion coefficient given in Eq. \ref{Dv} and $\zeta(p,T)$ denotes the the mean free path of H$_2$O molecules in air given by \cite{Pruppacher1997}:
\begin{equation}
    \zeta(p,T)=6.15\times 10^{-6}\bigg(\frac{1013.25}{p}\bigg)\bigg(\frac{T}{273.15}\bigg) \,.
\end{equation}
In this formulation, we neglect contributions involving radiative heating or cooling of the crystal’s surface, which are predominantly significant only at very small supersaturation levels (less than a few percent) \cite{Solch2010}.

Together, the equations for $\alpha^{(h)}$ and $\varphi^{(h)}$ form a coupled stochastic system that is written in a generic way as:
\begin{equation}
\left\{
\begin{aligned}
\frac{d\alpha^{(h)}}{dt} &= D_{\alpha}^{(h)}(\alpha,\varphi)\,, \\
\frac{d\varphi^{(h)}}{dt} &= D_{\varphi}^{(h)}(\alpha,\varphi) + \xi_\varphi(t)\,,
\end{aligned}
\right.\label{FPalphaphi}
\end{equation}
in terms of the drift coefficients $D_i$ ($i,j =\alpha,\varphi$) and the stochastic noise $\xi_i(t)$ with the conditions:
\begin{equation}
\left\langle \xi_i(t) \right\rangle^{(h)} = 0, \qquad 
\left\langle \xi_i(t), \xi_j(t') \right\rangle^{(h)} = 2\mathbb{D}_{ij}^{(h)}\delta(t - t'), \quad i,j = \alpha, \varphi\,,   \label{diffnoise}
\end{equation}
where $\xi_\alpha(t)$ is assumed zero from Eq. \eqref{eq1_effsurface}.
The matrix expressions for the drift  $D_{i}^{(h)}$  and diffusion   $\mathbb{D}_{ij}^{(h)}$ terms in Eqs. \eqref{FPalphaphi} and \eqref{diffnoise} are:
\begin{equation}\label{eqD1}
 \textbf{D}^{(h) }= 
  \left(
\begin{array}{c}
D_{\alpha}^{(h)}   \\
D_{\varphi }^{(h)} \\
\end{array}
\right)=
 \left(
\begin{array}{c}
\tfrac{2C^{(h)}_{0}A}{r_{0}^{(h)2}} \ \varphi \\
- \tfrac{4\pi N C_0^{(h)}r^{(h)}A  \rho}{\rho_a} \bigg(\tfrac{pR_v}{p_s(T)R_a}+\tfrac{L^2}{c_pR_v T^2}\bigg) \ \varphi\\
\end{array}
\right)\,,
\end{equation}
and
\begin{equation}\label{eqD2} 
\mathbb{D}^{(h)}= 
  \left(
\begin{array}{cc}
0 & 0 \\
0 & \mathbb{D}_{\varphi \varphi}^{(h)} \\
\end{array}
\right)=
 \left(
\begin{array}{cc}
0  & 0 \\
0 &  \frac{D'_{vh}}{\frac{D'_{vh}L p_s(T)}{k'_{h}T}\big(\frac{L}{R_vT}-1\big) +R_vT} \\
\end{array}
\right)\,.
 \end{equation} 
To the system described in Eq.~\eqref{FPalphaphi} is associated a general Fokker--Planck formulation (as in Eq.~\eqref{FP}) in terms of the probability distribution function $P_h(\alpha, \varphi, t)$, which describes the probability of finding a crystal of habit $h$ with effective area $\alpha$ and supersaturation $\varphi$ at time $t$:
\begin{align}
\partial_t P_h &= - \ D_\alpha^{(h)}\partial_\alpha   P_h  
-\partial_\varphi \left( D_\varphi^{(h)} P_h \right) + \mathbb{D}_{\varphi\varphi}^{(h)} \partial^2_\varphi  \  P_h \,, 
\end{align}
As seen in Section \ref{sec2}, this equation can be associated with a U(1) symmetry for individual habits. To extend this equation to a system of multiple, interrelated habits, we can use the results obtained for the SU(2) $\otimes$ U(1) group in order to generalize it into the compact form of new FP system equations for the four habits:
\begin{equation}\label{newFP}
     \partial_t P_h = - \ D_\alpha^{(h)} \partial_\alpha   P_h  
-\partial_\varphi \left( D_\varphi^{(h)} P_h \right) + \mathbb{D}_{\varphi\varphi}^{(h)}\partial^2_\varphi \  P_h   +  g_H I^{(1)}_h + g_H^2 I^{(2)}_{h}\,,\qquad h=1,...,4\,.
\end{equation}
The key innovation introduced by Eq.~\eqref{newFP} lies in the presence of the two additional integral operator terms $ I^{(1)}_h $ and $ I^{(2)}_h $, which are linear and quadratic in the coupling constant $g_H$, respectively \footnote{The integral operator terms $ I^{(i)}_a $ are defined in Section \ref{Nonabel} and in Appendix \ref{AppendixB}}. These terms encapsulate the mutual influence between the probability distributions of different crystal habits, and are defined as:
\begin{equation}
I^{(1,2)}_h(\alpha,\varphi,t) = 
\begin{cases}
    S^{(1,2)}_{a}(\varphi,t) + F^{(1,2)}_{a}(\alpha,t) + G^{(1,2)}_{a}(\alpha,\varphi) & \quad h=a=1,2,3\,\,, \\
    0 & \quad h=4\,\,.
\end{cases}
\end{equation}
The explicit forms of the $S^{(1,2)}_{a}$, $F^{(1,2)}_{a}$, and $G^{(1,2)}_{a}$ terms are detailed below. These expressions involve cross-integrals of the probability functions and their derivatives and reflect the nontrivial correlations arising among the different habits of the system.
\begin{equation}
\begin{aligned}
S^{(1)}_{a}(\varphi,t)&= \epsilon_{abc}  \int   d\alpha  \Big[ P^c(\partial_{\alpha} P_b)  - (\partial_t P_b) (D_{{\alpha}}^{(c)}P_{{c}}) +(D_{{j}}^{(c)}P_{{c}} -  \mathbb{D}_{jk}^{(c)}\partial_k   P_{c}) \partial_\alpha (D_{{j}}^{(b)} P_{{b}} -  \mathbb{D}_{jn}^{(b)}\partial_n   P_{b})    \\
& - \left(\partial_j (D_{{j}}^{(b)}P_{{b}} -  \mathbb{D}_{jn}^{(b)}\partial_nP_{b})    \right) (D_{{\alpha}}^{(c)}P_{{c}}) \Big]\\
F^{(1)}_{a}(\alpha,t)&= \epsilon_{abc} \int   d\varphi \,\Big[ P_c(\partial_\varphi P_b)  - (\partial_t P_b) (D_{{\varphi}}^{(c)}P_{{c}} -  \mathbb{D}_{\varphi\varphi}^{(c)}\partial_\varphi P_{c})(D_{{j}}^{(c)}P_{{c}} -  \mathbb{D}_{jk}^{(c)}\partial_k  P_{c}) \times  \\ 
\times &\left(\partial_\varphi (D_{{j}}^{(b)}P_{{b}} -  \mathbb{D}_{jn}^{(b)} \partial_nP_{b})\right) -\left(\partial_j (D_{{j}}^{(b)}P_{{b}} -  \mathbb{D}_{jn}^{(b)}\partial_n P_{b})    \right) (D_{{\varphi}}^{(c)}P_{{c}} -  D_{{\varphi\varphi}}^{(c)}\partial_\varphi P_{c}\Big] \\
G^{(1)}_{a}(\alpha,\varphi)&= \epsilon_{abc} \int dt \,  \left[  (D_{{i}}^{(c)}P_{{c}} -  \mathbb{D}_{ik}^{(c)}\partial_k  P_{c}(\partial_t (D_{{i}}^{(b)}P_{{b}} -  \mathbb{D}_{ij}^{(b)}\partial_jP_{b}))  - \left(\partial_i  (D_{{i}}^{(b)}P_{{b}} -  \mathbb{D}_{ij}^{(b)}\partial_j   P_{b})\right) P_c \right] 
\end{aligned}
\end{equation}
\begin{equation}
\begin{aligned}
S^{(2)}_a (\varphi,t) &= \int d\alpha \, \big( P_{b} P_a (D_{{\alpha}}^{(b)} -D_{{\alpha}}^{(a)}) \big)P_{b}  \\
F^{(2)}_{a}(\alpha,t)&=  \int {\rm d}\varphi \, \Big [P_b \big( P_a (D_{{\varphi}}^{(b)}P_{{b}} -  \mathbb{D}_{\varphi \varphi}^{(b)}\partial_\varphi  P_{b}) - P_b (D_{{\varphi}}^{(a)}P_{{a}} -  \mathbb{D}_{\varphi \varphi}^{(a)}\partial_\varphi  P_{a}) + (D_{{j}}^{(b)}P_{{b}} -  \mathbb{D}_{jk}^{(b)}\partial_k P_{b}) \ \times \\
\times &\Big((D_{{j}}^{(a)}P_{{a}} -  \mathbb{D}_{jm}^{(a)}\partial_m P_{a})  (D_{{\varphi}}^{(b)}P_{{b}} -  \mathbb{D}_{\varphi \varphi}^{(b)}\partial_\varphi  P_{b}) - (D_{{j}}^{(b)}P_{{b}} -  \mathbb{D}_{jm}^{(b)}\partial_m P_{b})  (D_{{\varphi}}^{(a)}P_{{a}} -  \mathbb{D}_{\varphi \varphi}^{(a)}\partial_\varphi  P_{a})\Big) \Big]\\
G^{(2)}_{a}(\alpha,\varphi)&=  \int {\rm d}t \  \Big[ (D_{{i}}^{(b)}P_{{b}} -  \mathbb{D}_{ik}^{(b)}\partial_k P_{b}) \big( P_b(D_{{i}}^{(a)}P_{{a}} -  \mathbb{D}_{im}^{(a)}\partial_m P_{a})  -  P_a(D_{{i}}^{(b)}P_{{b}} -  \mathbb{D}_{im}^{(b)}\partial_m P_{b}) \big) \Big]
\end{aligned}
\end{equation}
These new terms represent the novel contribution introduced in this model. The resulting system of equations is not analytically solvable and requires numerical simulations, which are not analyzed in this work. In this generalization of the equations, the coupling constant $g_H$ acts as a mediator of the "strength" of these new terms. 
As discussed in previous sections, when considering multiple habits, we can introduce a total probability function, denoted by $\mathcal{P}$, as the sum of the individual probabilities, subject to the normalisation condition:
\begin{equation}\label{norm}
     \int_{-1}^{+\infty}\int_0^{+\infty}\mathcal{P}(\alpha,\varphi,t) \, d\alpha d\varphi = \sum_{h=1}^4 \frac{1}{4}\int_{-1}^{+\infty}\int_0^{+\infty}P_h(\alpha,\varphi,t)\,d\alpha d\varphi =1\,\,,
\end{equation}
and that it will be preserved over time:
\begin{equation}
    \frac{d}{dt}\sum_{h=1}^4 \int_{-1}^{+\infty}\int_0^{+\infty}P_h(\alpha,\varphi,t)d\alpha d\varphi=  0 \,\,,
\end{equation}
This ensures that the probabilistic structure of the model is internally consistent and conserved.

\section{Conclusions}\label{sec5}
In this paper, we propose a novel theoretical framework for modelling the growth dynamics of ice crystal habits in ice clouds, focusing on the depositional regime. Using tools from classical field theory and gauge symmetry, we developed a heuristic extension of the standard Fokker–Planck formalism to describe the evolution and interaction of the four primary ice crystal habits commonly used to model the ice clouds microphysics: droxtals, plates, columns, and rosettes.

Starting from a field theory framework with a U(1) gauge symmetry, which allows the derivation of the standard Fokker-Planck equation, we heuristically extended the formalism to a non-Abelian gauge symmetry described by the group SU(2) $\otimes$ U(1). This extension leads to a system of coupled Fokker-Planck equations that simultaneously govern the growth dynamics of the four primary crystal habits. In this model, the SU(2) sector governs the interactions and transitions among more complex habits (columns, plates, and rosettes) while the U(1) component describes droxtals, which dominate early-stage nucleation and growth under very low-temperature conditions. 

At the current stage, the droxtal field is dynamically decoupled from the other components, and the full SU(2) $\otimes$ U(1) symmetry remains preserved. However, a natural extension of the framework involves the introduction of interaction terms between the U(1) and SU(2) subspaces. These terms would enable continuous transitions among all crystal habits, effectively modelling the evolution from initial nucleation to more developed morphologies. Such interactions would explicitly break the original gauge symmetry, which would then survive only as a residual structure under specific thermodynamic conditions. The detailed formulation of these symmetry-breaking mechanisms is left for future work.

A key feature of this theoretical construction is the emergence of a new source term in the coupled Fokker–Planck equations. This term governs transitions within the SU(2) subspace and encapsulates the influence of habit interactions on the overall growth dynamics. The solutions to these equations describe the temporal evolution of probability distributions for each crystal habit, as functions of relevant atmospheric parameters such as supersaturation and the effective surface area of the ice crystals.

An essential next step will be the numerical resolution of the proposed system of coupled stochastic equations, which do not admit closed-form analytical solutions. Numerical simulations will be key for comparing model predictions with in-situ and satellite observations of ice crystal habit distributions, as well as for estimating the coupling constants $g_H$ associated with habit interactions. The study of the constant coupling represents a further future objective, with the aim of quantifying it and assessing the potential existence of different thermodynamic regimes. 

Ultimately, this work outlines a new path for embedding field-theoretic principles into the modelling of ice crystal growth. Given the central role of cirrus clouds in the Earth's radiation budget, covering over 30\% of the planet and representing a major source of uncertainty in current climate models, developing accurate theoretical tools to capture their micro-physical behavior is critical. By bridging the language of gauge theory with the structure of stochastic growth equations, this study lays the groundwork for a more physically predictive description of ice clouds within the broader context of climate modelling.

\section{Acknowledgements}
We would like to thank Germán Rodrigo and German Sborlini for their careful reading of the manuscript and for their valuable comments. This work is supported by the Spanish Government (Agencia Estatal de Investigación MCIN/ AEI/10.13039/501100011033) Grants No. PID2020-114473GB-I00, No. PID2022-141910NB-I00, No. PID2023-146220NB-I00 and No. CEX2023-001292-S. LC is supported by Generalitat Valenciana GenT Excellence Programme (CIDEGENT /2020/011). Part of the research activities described in this paper were carried out with contribution of the Next Generation EU funds within the National Recovery and Resilience Plan (PNRR), Mission 4 - Education and Research, Component 2 - From Research to Business (M4C2), Investment Line 3.1 - Strengthening and creation of Research Infrastructures, Project IR0000038 – “Earth Moon Mars (EMM)”. EMM is led by INAF in partnership with ASI and CNR. The work is supported by the FIT-FORUM project CUP: F33C23000240005 and CASIA project  CUP: F93C23000430001 of the ASI.

\newpage
\appendix
\section{A Primer on Gauge Symmetries for Habit Dynamics}
\label{sec:Gaugegroups}
\setlength{\parindent}{0pt}
The purpose of this Appendix is to provide a brief, self-contained introduction to the concepts of gauge symmetry and Lie groups \cite{georgi1999lie}, specifically tailored to their application in our model of ice crystal habit evolution. This Appendix is intended as a primer for readers less familiar with the topic; readers expert in quantum field theory may wish to skip this section.

The generators of the special unitary group SU(N), $T_a$ satisfy the Lie algebra commutation relations:
\begin{equation}
   [T_a, T_b] = i f_{abc} T_c\,, \qquad f_{abc} = -f_{bac}\,,
\end{equation}
where $f_{abc}$ are the antisymmetric structure constants of the group. It is a Lie group of dimension $\rm{N}^2 - 1$, which corresponds to the number of independent generators required to define any transformation within the group. In the particular case of SU(2) the generators of the group are the well known Pauli matrices $\sigma^a$ ($T^a=\sigma^a/2$, with $a=,1,2,3$) and the structure constants are the completely anti-symmetric Levi-Civita tensors $\epsilon_{abc}$ (\textit{i.e} $f_{abc} = \epsilon_{abc}$).
 
More clearly, an element $U$ of SU(N) in a given representation can be written as:
\begin{equation}
   U = \exp \left( i \sum_{a=1}^{N^2 - 1} \theta_a T_a \right)\,,
\end{equation}
where the $T_a$ are the generators of the group. The generic element $U$ could be, for instance, the generator of a local gauge transformation as defined in (and below) Eq.~\eqref{AtransfSU2}. Notice that for the sake of simplicity in the text, we omitted the summation symbol for repeated indices. In the fundamental representation, the $T_a$ are $\rm N \times N$ Hermitian matrices:
\begin{equation}
   T_a^\dagger = T_a, \qquad \text{Tr}(T_a) = 0\,.
\end{equation}
The traceless condition follows from the constraint $\det U = 1$. The generators satisfy the normalization condition also:
\begin{equation}
   \text{Tr}(T_a T_b) = \frac{1}{2} \delta_{ab}\,.
\end{equation}
Additionally, the generators satisfy the Jacobi identity:
\begin{equation}
   [T_a, [T_b, T_c]] + [T_b, [T_c, T_a]] + [T_c, [T_a, T_b]] = 0\,,
\end{equation}
which implies the following relation for the structure constants:
\begin{equation}
   f_{abe} f_{cde} + f_{bce} f_{ade} + f_{cae} f_{bde} = 0\,.
\end{equation}
In particular, for N=2 (\textit{i.e} for SU(2)) the following explicit relation holds for the structure constants
\begin{equation}
    \epsilon_{abc} \, \epsilon_{blm} = \delta_{am} \delta_{cl} - \delta_{al} \delta_{cm}\,.
\end{equation} 
In the adjoint representation of SU(N), the generators are represented as $\rm (N^2 - 1) \times (N^2 - 1)$ matrices:
\begin{equation}
   \left(T^{(\text{adj})}_a\right)_{bc} = -i f_{abc}\,.
\end{equation}
This representation acts naturally on the Lie algebra itself.

\section{Explicit calculation of the integral operators} \label{AppendixB}
In section \ref{Nonabel} we introduced the integral operators $\mathcal{I}^{(i)}$ which denote the interaction or \textit{mixing} between the different habits. Here we report on the explicit calculation when the actual form of the $A_\mu$ fields (see Eq.~\eqref{Amu}) is taken into account. In order to proceed, it is more convenient to take the alternative definition for these integral operators:
\begin{equation}
\label{eq:Itilde_vsII}
I^{(i)}_a = f_{abc}\, \mathcal{I}^{(i)}(\mathbb{A}^b,\mathbb{A}^c)\,\,,\qquad \,\, i = 1,2\,.
\end{equation}
Therefore the explicit expression of the integral operators in term of the $A_\mu$ fields is
\begin{align}
    I^{(1)}_a & = f_{abc}\int {\rm d}x^\nu  \left[A_\mu^c\,\left( \partial^\nu A^\mu_b - \partial^\mu A^\nu_b \right) - \partial_\mu\left( A^\mu_b A^\nu_c\right)\right] \,\,,\\
    I^{(2)}_a & = f_{abc} \int {\rm d}x^\nu  f_{blm}\, A^\mu_m \, A^\nu_l \, A_\mu^c \,\,.
\end{align}
We start discussing the first order integral operator $I^{(1)}_a$ which takes the explicit form: 
\begin{equation}
\begin{aligned}
    I^{(1)}_a    &=  f_{abc}\int {\rm d}x^\nu   \left[  A^c_\mu(\partial^\nu A^\mu_b)  - (\partial_\mu A^\mu_b) A^\nu_c \right]\,.
\end{aligned}    
\end{equation}
We then split into spatial and temporal components as:
\begin{equation}
    I^{(1)}_a =  f_{abc}\int {\rm d}x^i \left[  A^c_\mu(\partial^i A^\mu_b)  - (\partial_\mu A^\mu_b) A^i_c \right] +  f_{abc}\int {\rm d}t \left[  A^c_\mu(\partial_t A^\mu_b)  - (\partial_\mu A^\mu_b) A^0_c \right]\,.
\end{equation}
Writing explicitly the components  $\mu=0,i=1,2,3$ and we find four integrals:
\begin{equation}
\begin{aligned}
I^{(1)}_a =  & \underbrace{ f_{abc}\int {\rm d}x^i \left[  A^c_j(\partial^i A^j_b)  - (\partial_j A^j_b) A^i_c \right]}_{I^{(1)}_{I}} + 
 \underbrace{ f_{abc}\int {\rm d}x^i \left[  A^c_0(\partial^i A^0_b)  - (\partial_t A^0_b) A^i_c \right]}_{I^{(1)}_{II}} \\
 &+ \underbrace{ f_{abc}\int {\rm d}t \left[  A^c_i(\partial_t A^i_b)  - (\partial_i A^i_b) A^0_c \right]}_{I^{(1)}_{III}} + \underbrace{ f_{abc}\int {\rm d}t \left[  A^c_0(\partial_t A^0_b)  - (\partial_t A^0_b) A^0_c \right]}_{I^{(1)}_{IV}}\,,
\end{aligned}
\end{equation}
where the integral $ I^{(1)}_{IV}$ vanishes. The remaining integrals are explicitly given below when we consider the explicit expression for the $A_\mu$ fields given in Eq.~\eqref{Amu},
\begin{equation}
\begin{aligned}
I^{(1)}_{I} &=  f_{abc} \int {\rm d}x^i  \Big[  (D^{{j}}_{(c)}P_{{c}} -  \mathbb{D}^{jk}_{(c)}\partial_k P_{c}) \left(\partial^i (D^{{j}}_{(b)}P_{{b}} -  \mathbb{D}^{jn}_{(b)}\partial_n  P_{b})\right)  \\
& \qquad \qquad - \left(\partial_j (D^{{j}}_{(b)}P_{{b}} -  \mathbb{D}^{jn}_{(b)}\partial_n  P_{b})    \right) (D^{{i}}_{(c)}P_{{c}} -  \mathbb{D}^{ik}_{(c)}\partial_k  P_{c}) \Big]\,,
\end{aligned}
\end{equation}
\begin{equation}
\begin{aligned}
I^{(1)}_{II} &= f_{abc} \int {\rm d}x^i  \left[ P^c(\partial^i P_b)  - (\partial_t P_b) (D^{{i}}_{(c)}P_{{c}} -  \mathbb{D}^{ik}_{(c)}\partial_k  P_{c})\right]    \,,
\end{aligned}
\end{equation}
\begin{equation}
\begin{aligned}
I^{(1)}_{III} &= f_{abc} \int {\rm d}t   \left[  (D^{{i}}_{(c)}P_{{c}} -  \mathbb{D}^{ik}_{(c)}\partial_k P_{c})(\partial_t (D^{{i}}_{(b)}P_{{b}} -  \mathbb{D}^{ij}_{(b)} \partial_j P_{b}))  - \left(\partial_i  (D^{{i}}_{(b)}P_{{b}} -  \mathbb{D}^{ij}_{(b)}\partial_j   P_{b})\right) P_c \right]\,.    
\end{aligned}
\end{equation}
The expression of the second-order integral operator $I^{(2)}_a$ cannot be further simplified in the general SU(N) case (see Appendix \ref{sec:Gaugegroups}). However, since in Sections~\ref{sec3} and~\ref{sec4} the specific case N $ = 2$ is considered (more precisely, SU(2)~$\otimes$~U(1)), we report in the final part of this Appendix only results valid when the SU(2) group is taken into account (note that for the integral $I^{(1)}_a$ it is sufficient to impose $f_{abc} = \epsilon_{abc}$). In this case, we can exploit the useful identity $\epsilon_{abc} \, \epsilon_{blm} = \delta_{am} \delta_{cl} - \delta_{al} \delta_{cm}$ (defined in Appendix \ref{sec:Gaugegroups}), which simplifies the integral $I^{(2)}_a$ as follows:
\begin{equation}\label{Jab}
 I^{(2)}_a=  - \epsilon_{abc} \int {\rm d}x^\nu  \epsilon_{blm}\, A^\mu_m \, A^\nu_l \, A_\mu^c = \int  A_{\mu b}(A_a^{\mu} A^{\nu b} - A^{\mu b} A_a^{\nu})\, dx_\nu \,.
\end{equation}
Note that we have considered the sum only for $b \neq a$ because for $b=a$ the integrals goes to zero.
Again we can split the integral into the spatial and temporal components:
\begin{equation}
 \ \int  A_{\mu b}(A_a^{\mu} A^{i b} - A^{\mu b} A_a^{i})\, dx_i + \int  A_{\mu b}(A_a^{\mu} A^{0 b} - A^{\mu b} A_a^{0})\, dt\,.
\end{equation}  
We can explicit the index $\mu=0,i=1,2,3$ and we find four integrals:
\begin{equation}
\begin{aligned}
  & \underbrace{\int  \left[A_{j b}(A_a^{j} A^{i b} - A^{j b} A_a^{i}) \right] \, dx^i}_{I^{(2)}_{I}} + 
 \underbrace{\int \left[ A_{0 b}(A_a^{0} A^{i b} - A^{0 b} A_a^{i})\right] \, dx^i}_{I^{(2)}_{II}} \\
 &+ \underbrace{\int  \left[ A_{i b}(A_a^{i} A^{0 b} - A^{i b} A_a^{0}) \right]\, \mathrm{d}t}_{I^{(2)}_{III}} + \underbrace{\int  \left[ A_{0 b}(A_a^{0} A^{0 b} - A^{0 b} A_a^{0})\right] \, \mathrm{d}t}_{I^{(2)}_{IV}}\,.
\end{aligned}
\end{equation}
The integrals $ I^{(2)}_{IV}$ is zero because the term in parentheses vanishes. The other integrals, after several calculations, are explicitly given below.
\begin{equation}\label{I2termI}
\begin{aligned}
I^{(2)}_{I} &=  \int {\rm d}x^i  \,  \Big[ (D^{{j}}_{(b)}P_{{b}} -  \mathbb{D}^{jk}_{(b)}\partial_k P_{b}) \big((D^{{j}}_{(a)}P_{{a}} -  \mathbb{D}^{jm}_{(a)}\partial_m P_{a})  (D^{{i}}_{(b)}P_{{b}} -  \mathbb{D}^{in}_{(b)}\partial_n  P_{b})\\
 & \qquad - (D^{{j}}_{(b)}P_{{b}} -  \mathbb{D}^{jm}_{(b)}\partial_m P_{b})  (D^{{i}}_{(a)}P_{{a}} -  \mathbb{D}^{in}_{(a)}\partial_n  P_{a})\big) \Big]\,,
\end{aligned}
\end{equation}
\begin{equation}\label{I2termII}
\begin{aligned}
   I^{(2)}_{II}= \int {\rm d}x^i \, \Big [P_b \big( P_a (D^{{i}}_{(b)}P_{{b}} -  \mathbb{D}^{ik}_{(b)}\partial_k  P_{b}) - P_b (D^{{i}}_{(a)}P_{{a}} -  \mathbb{D}^{ik}_{(a)}\partial_k  P_{a}) \Big]\,,
\end{aligned}
\end{equation}
\begin{equation}\label{I2termIII}
\begin{aligned}
I^{(2)}_{III} &=  \int {\rm d}t \  \Big[ (D^{{i}}_{(b)}P_{{b}} -  \mathbb{D}^{ik}_{(b)}\partial_k P_{b}) \big( P_b(D^{{i}}_{(a)}P_{{a}} -  \mathbb{D}^{im}_{(a)}\partial_m P_{a})  -  P_a(D^{{i}}_{(b)}P_{{b}} -  \mathbb{D}^{im}_{(b)}\partial_m P_{b}) \big) \Big]\,.
\end{aligned}
\end{equation}

\section{Noether current and conserved habit charge}
\label{AppendixC}
The Noether current associated to a global symmetry for the SU(N) theory defined in section \ref{sec3} can be obtained as follows:
\begin{equation}
j^{\mu, a} = \frac{\partial \mathcal{L}}{\partial(\partial_\mu A_\nu^b)} \delta^{(a)} A_\nu^b = f^{a b c} F^{\mu\nu, b} A_\nu^c\,,
\end{equation}
with $f^{a b c}$ the SU(N) structure constants. In the U(1) case, the structure constants are vanishing and therefore the associated charge is identically zero. 

In the specific case of the SU(2) group we have:
\begin{equation}
j^{\mu, a} = \frac{\partial \mathcal{L}}{\partial(\partial_\mu A_\nu^b)} \delta^{(a)} A_\nu^b = \epsilon^{a b c} F^{\mu\nu, b} A_\nu^c\,,
\end{equation}
and the corresponding charges are obtained by the three dimensional integral of the temporal component of $j^{\mu, a}$:
\begin{equation}
\begin{aligned}
Q_a &= \int {\rm d}^3x \, j^{0}_a = \int {\rm d}^3x \, \epsilon_{a b c} F^{0i}_b A^i_c = \int {\rm d}^3x \, \epsilon_{a b c} \left( \partial^0 A^{i}_b - \partial^i A^{0}_b - g_H \epsilon_{b d e} A^{0}_d A^{i}_e \right) A^i_c \\
&= \int {\rm d}^3x \, \epsilon_{a b c} \left( \partial^0 \left(D^i_{{(b)}}P_b   -  \mathbb{D}_{(b)}^{ij}\partial_jP_{b} \right)  - \partial^i P_b \right) \left(D^i_{{(c)}}P_c   -  \mathbb{D}_{(c)}^{ij}\partial_j  P_{c} \right) +\\
&- g_H \int {\rm d}^3x \, \left(  P_a \left(D^i_{{(c)}}P_c   -  \mathbb{D}_{(c)}^{ij}\partial_j   P_{c})\right)^2  -  \left(D^i_{{(a)}}P_a   -  \mathbb{D}_{(a)}^{ij}\partial_j  P_{a} \right) P_c \left(D^i_{{(c)}}P_c   -  \mathbb{D}_{(c)}^{ij}\partial_j  P_{c} \right)\right)\,.
\end{aligned}
\end{equation}
A physical interpretation of the concept of habit charge within our model could be the following. The habit charge associated with the three SU(2) generators quantifies the tendency of a given habit to transform into another. The higher the habit charge, the greater the probability for that habit to change morphology. Its conservation imposes a ``selection rule'' on which habits can interact and transform into each other.

In contrast, the shape associated to U(1) is characterized by a vanishing habit charge. This implies that the corresponding habit does not spontaneously evolve into other forms, unless a mixing term is explicitly introduced in the Lagrangian, as discussed in Section~\ref{sec3}.\\

\newpage

\bibliographystyle{abbrv}
\bibliography{references_new.bib}

\end{document}